\documentstyle[12pt]{article}
\newcommand{\proof}[1]{{\bf Proof:}
#1~$\Box$\vspace{.30cm}}
\newcommand{\bb}{\begin{equation}}
\newcommand{\ee}{\end{equation}}
\newcommand{\bqn}{\begin{eqnarray}}
\newcommand{\eqn}{\end{eqnarray}}

\newcommand{\tB}{\tilde B }

\newtheorem{theorem}{Theorem}

\newtheorem{lemma}{Lemma}

\begin{document}

\begin{titlepage}

\begin{flushright}

ULB-TH-99/30\\
DAMTP-1999-172 \\
hep-th/9912052

\end{flushright}

\begin{center}
{\Large {\bf A Theorem on First-Order Interaction
Vertices for Free $p$-Form 
Gauge Fields}}

\end{center}
\vfill

\begin{center}
{\large
Marc Henneaux$^{a,b}$ and Bernard Knaepen$^{a,c}$ 
\footnote{henneaux@ulb.ac.be, bknaepen@ulb.ac.be}}
\end{center}
\vfill

\begin{center}{\sl
$^a$ Physique Th\'eorique et Math\'ematique, Universit\'e Libre de
Bruxelles,\\
Campus Plaine C.P. 231, B--1050 Bruxelles, Belgium\\[1.5ex]

$^b$ Centro de Estudios Cient\'\i ficos de Santiago,\\
Casilla 16443, Santiago 9, Chile\\[1.5ex]

$^c$ DAMTP, Silver Street, Cambridge CB3 9EW, UK
}\end{center}

\vfill

\begin{abstract}
The complete proof of a theorem announced in
\cite{HK1} on the consistent interactions
for (non-chiral)
exterior form gauge fields is given.
The theorem can be easily generalized to the 
analysis of anomalies.  Its
proof amounts to computing the local BRST
cohomology $H^{0} (s\vert d)$ in the space of local $n$-forms
depending on the fields, the ghosts, the antifields
and their derivatives.  

\end{abstract}
\vfill
\end{titlepage}

\section{Introduction}
\setcounter{equation}{0}
\setcounter{theorem}{0}

Geometric attempts to generalize the Yang-Mills construction to $p$-form
gauge fields with $p>1$ 
have led to no-go results that indicate that this goal
cannot be achieved while maintaining spacetime locality 
\cite{Nepomechie,Teitelboim1,DaDeCa1}.

In fact, self-interactions of $p$-form gauge fields are so 
constrained that one can completely list them, even if one drops any a
priori geometric interpretation of the $p$-forms as connections for 
extended objects.  This task was
explicitly performed in  
\cite{HK1}, where
the following question was analyzed.
Consider the free action, 
\bb
I =\int d^n x\sum_a \big({-1 \over 2(p_{a} +1)!}H^a_{\mu_1
\ldots
\mu_{p_{a}+1}} H^{a \mu_1
\ldots \mu_{p_{a}+1}}\big), \label{Lagrangian}
\ee
for a system
of (non-chiral) exterior form gauge fields 
$B^a_{\mu_1 \ldots \mu_{p_{a}}}$ of degree
$\geq 2$.
Here, the $H^a$'s are the ``field strengths"
or ``curvatures",
\begin{eqnarray}
H^a&=&{1\over (p_{a}+1)!}
H^a_{\mu_1 \ldots \mu_{p_{a}+1}}dx^{\mu_1} \ldots
dx^{\mu_{p_{a}+1}} =dB^a, \label{FieldStrength}\\
B^a&=&{1\over p_{a}!} B^a_{\mu_1 \ldots \mu_{p_{a}}}
dx^{\mu_1} \ldots dx^{\mu_{p_a}}.
\end{eqnarray}
We assume throughout that  
the spacetime dimension satisfies
the condition $n>p_a+1$ for each
$a$ so that all the $p_a$-forms have local degrees of
freedom.
The action (\ref{Lagrangian}) is invariant under
the abelian gauge transformations,
\begin{equation}
B^a \rightarrow B^a + d \Lambda^a,
\label{origaugetr}
\end{equation}
where $\Lambda^a$ are arbitrary $p_a-1$ forms.  The
equations of motion, obtained by varying the fields
$B^a_{\mu_1 \ldots \mu_{p_a}}$, are given by,
\begin{equation}
\partial_\rho H^{a \rho \mu_1 \ldots \mu_{p_a}}=0
\Leftrightarrow d \overline{H}^a=0,
\end{equation}
where $\overline{H}^a$ is the dual of $H^{a \rho \mu_1
\ldots \mu_{p_a}}$.

The question addressed in \cite{HK1} was: what are the consistent (local)
interactions that can be added to the free action (\ref{Lagrangian})?
Interaction terms are said to be consistent if their preserve the
number (but not necessarily the form) of the independent
gauge symmetries.

Of course, one can always add to (\ref{Lagrangian})
gauge-invariant interaction terms constructed out of the curvature
components and their derivatives,
\bb
\int f(H^{(k)}_{\mu_1 \dots \mu_{p_k +1}},
\partial_\nu H^{(k)}_{\mu_1 \dots \mu_{p_k +1}}, \cdots,
\partial_{\nu_{1} \dots \nu_q} H^{(k)}_{\mu_1 \dots \mu_{p_k +1}}) d^nx.
\ee
Being strictly gauge-invariant, these terms actually do not deform
the gauge symmetries.  One may, however, also
search for interaction terms that deform not only the
action, but also the gauge transformations.
These turn out to be extremely scarce, as the
following theorem indicates:
\begin{theorem}
Besides the obvious gauge-invariant
interactions, 
the only consistent interaction vertices 
that can be added to (\ref{Lagrangian}) have the Noether form,
\bb
V = \sum_{(A)} g_{(A)} V_{(A)}
\label{vertex0}
\ee
where the $g_{(A)}$ are the coupling constants and the $V_{(A)}$
read
\bb
V_{(A)} = \int j^{(t)} \wedge B^{(t)}.
\label{vertex}
\ee
Here, $j^{(t)}$ are gauge-invariant conserved $(n-p_t)$-forms,
$dj^{(t)} \approx 0$,  and therefore, are exhausted by the exterior
polynomials in the curvature forms $H^{(k)}$
and their duals $\bar{H}^{(k)}$ {\em \cite{HKS1}}.
\label{central}
\end{theorem}
Because $j^{(t)}$ must have exactly form-degree $n-p_t$,
so that the form degree of the integrand of (\ref{vertex})
matches the spacetime dimension $n$, there may be no vertex
of the type (\ref{vertex0}) for given spacetime dimension and
form-degrees of the exterior form gauge fields.  For example,
a set of $2$-form gauge fields admits gauge symmetry-deforming
non-trivial interactions only
in $n=4$ dimensions
\cite{HenneauxPL2} and these are of the Freedman-Townsend type
\cite{FT1}.  Other examples of vertices
of the form (\ref{vertex}) involving $p$-form
gauge fields of different form degrees are provided by the Chapline-Manton
interactions
\cite{ChaplineManton,Nito,Cham1,Cham2,BergRooWitNieu,Baulieu1}. The
analysis of
\cite{HK1} also enabled one to exhibit new
symmetry-deforming interactions, but again only in special dimensions
(see also \cite{BrandtD1}; these 
interactions have been further analysed in \cite{BrandtT1,BrandtT2}).

In (\ref{vertex}), the $j^{(t)}$ are exterior polynomials in $H^{(k)}$
and $\bar{H}^{(k)}$ with coefficients that can involve $dx^\mu$.
If one imposes Lorentz invariance, bare $dx^\mu$'s cannot appear. Note
also that if $(n-1)$-forms are included, an infinite number of couplings
(\ref{vertex}) may in general be constructed since arbitrary powers of
the duals (which are zero forms) can appear.

The vertices (\ref{vertex0}) have a number of remarkable
properties:
\begin{enumerate}
\item First, while the strictly gauge-invariant vertices may involve
derivatives of the individual components
$H^{(k)}_{\mu_1 \dots \mu_{p_k +1}}$ of the curvatures,
the vertices (\ref{vertex}) are very special: they
can be expressed as polynomials
in the exterior product (``exterior polynomials") in the
(undifferentiated) forms $B^{(k)}$, $H^{(k)}$ and  $\bar{H}^{(k)}$.
This is not an extra requirement. Rather, this
property follows directly from the demand that
(\ref{vertex0}) defines a consistent interaction.
\item If the vertices (\ref{vertex0}) do not involve the duals
$\bar{H}^{(k)}$, one
recovers the familiar Chern-Simons terms \cite{DJT1}.  These are
off-shell gauge-invariant up to a total
derivative and so, do not deform the gauge transformations.
Vertices (\ref{vertex0}) involving the duals are only on-shell
gauge-invariant up to a total derivative.  These vertices
do deform the
gauge transformations.
\item Although the vertices (\ref{vertex0}) deform the
gauge symmetries when they involve the duals $\bar{H}^{(k)}$,
they do not modify the algebra of the gauge transformations
(to the first order in the coupling constants considered here)
because they are linear in the $p$-form
potentials.  This is in sharp contrast with the
Yang-Mills construction, which yields a vertex of the form
$\bar{H}^a \wedge B^b \wedge B^c$. There is thus no room for an analog of the 
Yang-Mills vertex for exterior forms of degree $\geq 2$.
How the result is amended in the
presence of $1$-forms will be discussed
at the end.
\item The fact that the gauge transformations remain
abelian to first-order in the coupling constant is not in contradiction
with \cite{LavLuPoSt1}. Indeed, we focus here only on symmetries of
the equations of motion that are also symmetries of the action.
Furthermore, the non-abelian structure uncovered in \cite{LavLuPoSt1}
concerns symmetries associated with non-trivial global
features of the spacetime manifold, which are rigid
symmetries \cite{CrJuLuPo1}.
\end{enumerate}

The above theorem was stated and discussed in \cite{HK1}
but a complete demonstration of it was not given.  The purpose
of this paper is to fill this gap.  As we shall see, the proof has an interest
in itself since it illustrates various cohomologies arising in local
field theory.

We conclude this introduction by observing
that the interaction vertices are in general not duality-invariant, in the
sense that an interaction vertex that
is available in one version of the theory
may not be so in the dual version where some of the
$p$-form potentials are traded
for ``dual" $(n-p-2)$-form potentials.

\section{Consistent interactions and Local BRST Cohomology}
\setcounter{equation}{0}
\setcounter{theorem}{0}

Our approach to the problem of constructing consistent interaction
vertices for a gauge theory is based on the BRST symmetry.
As shown in \cite{BH1,HenneauxCont1}, the question boils down
to computing the local BRST cohomological group at ghost
number zero in the algebra of local $n$-forms depending on the
fields, the ghosts, the antifields and their derivatives.
These groups are denoted by $H^0(s\vert d)$.  The cocycle condition
reads,
\bb
sa + db =0,
\label{cocycle}
\ee
where $a$ (respectively $b$)
is a local $n$-form (respectively $(n-1)$-form)
of ghost number zero (respectively one).  Trivial
solutions of (\ref{cocycle}) are of the form,
\bb
a = s m + dn
\label{coboundary}
\ee
where $m$ (respectively $n$) is a local $n$-form
(respectively, $(n-1)$-form) of ghost number
$-1$ (respectively $0$).  One often refers to (\ref{cocycle})
as the ``Wess-Zumino consistency condition" \cite{WZ}.

If $a$ is a solution of (\ref{cocycle}), its
antifield-independent part defines a consistent interaction;
and conversely, given a consistent interaction, one can complete
it it by antifield-dependent terms to get a BRST cocycle
(\ref{cocycle}).  As explained in \cite{BH1,HenneauxCont1},
it is necessary to include the antifields in the analysis
of the cohomology in order to cover symmetry-deforming
interactions.

In the case at hand, the gauge symmetries are
reducible and the following set of antifields is required
\cite{BV2,HenneauxTeitelboim},
\begin{equation} B^{*a \mu_1 \ldots \mu_{p_a}}, B^{*a\mu_1
\ldots \mu_{p_a-1}},\ldots, B^{*a\mu_1},B^{*a}.
\label{antifieldlist}
\end{equation}
The Grassmann parity and the {\it antighost}
number of  the antifields $B^{*a \mu_1
\ldots \mu_{p_a}}$ associated with the fields $B^a_{\mu_1
\ldots \mu_{p_a}}$ are equal to $1$.  The Grassmann parity
and the {\it antighost} number of the other antifields is
determined according to the following rule. As one moves
from one term to the next one to its right in
(\ref{antifieldlist}),  the Grassmann parity changes and
the antighost number increases by one unit. Therefore the
parity and the antighost number of a given antifield
$B^{*a \mu_1 \ldots \mu_{p_a-j}}$ are respectively $j+1$
modulo $2$ and $j+1$.

Reducibility also imposes the following set of
ghosts,
\begin{equation} C^a_{\mu_1 \ldots
\mu_{p_a-1}},\ldots,C^a_{\mu_1
\ldots
\mu_{p_a-j}},\ldots, C^a.
\label{ghosts}
\end{equation} 
These ghosts carry a degree called the pure
ghost number. The pure  ghost number of $C^a_{\mu_1
\ldots\mu_{p_a-1}}$ and its grassmann parity are equal to
1.   As one moves from one term to the next one to its right
in  (\ref{ghosts}), the Grassmann parity changes and the
ghost  number increases by one unit up to $p_{a}$.

We denote by ${\cal P}$ the algebra of spacetime
forms with coefficients that are polynomials in the fields,
antifields, ghosts and their derivatives.

The action of $s$ in ${\cal P}$ is the sum of two parts,
namely, the ``Koszul-Tate differential $\delta$" and the
``longitudinal exterior derivative $\gamma$":
\begin{equation}
s=\delta +\gamma,
\end{equation}
where we have,
\begin{eqnarray}
\delta B^a_{\mu_1 \ldots \mu_{p_a}}&=&0, \\
\delta C^a_{\mu_1 \ldots \mu_{p_a-j}}&=&0, \\
\delta {\overline B}^{*a}_1 +d{\overline H}^a &=&0,
\nonumber \\
\delta {\overline B}^{*a}_2 +d{\overline B}^{*a}_1
&=&0,\nonumber \\ &\vdots& \label{defduaux} \\
\delta {\overline B}^{*a}_{p_a+1}+d{\overline
B}^{*a}_{p_a} &=& 0,
\nonumber
\end{eqnarray}
and,
\begin{eqnarray}
\gamma {{B}^{*a\mu_1 \dots \mu_{p_a+1-j}}}&=&0,\\
\gamma B^a + dC^a_1 &=&0 ,\\
\gamma C^a_1 + dC^a_2 &=&0,\\
&\vdots& \nonumber \\
\gamma C^a_{p_a-1} + dC^a_{p_a} &=&0,\\
\gamma C^a_{p_a} & = &0.
\end{eqnarray}
In the above equations,
$C^a_{j}$ is the
$(p_a-j)$-form whose components are $C^a_{\mu_1 \ldots
\mu_{p_a-j}}$. Furthermore, we have systematically denoted
(as above) the duals by an
overline to avoid confusion with the *-notation of
the antifields. The actions of $\delta$ and $\gamma$ on the 
individual components of the antifields (\ref{antifieldlist}), 
ghosts (\ref{ghosts}) and their derivatives are easily 
read off from the above formulas (recalling that 
$\delta (dx^\mu)=\gamma (dx^\mu)=0$,
$[\partial_\mu,\delta]=0,[\partial_\mu,\gamma]=0$).

\section{General procedure for working out BRST cohomology}
\label{ggen}
\setcounter{equation}{0}
\setcounter{theorem}{0}
\setcounter{lemma}{0}

In order to prove the theorem, we shall solve the BRST cocycle
condition by proceeding as in the Yang-Mills case \cite{BH2,BBH2}.
To that end, one expands the cocycles and the cocycle
condition according to the antighost number.
Thus, if $a$ is a BRST cocycle (modulo $d$),
then its various components in the expansion,
\bb
a = a_0 + a_1 + a_2 + \cdots + a_k, \; \; antigh(a_i) = i,
\label{expansion}
\ee
must fulfill the chain of equations,
\begin{eqnarray}
\label{chain}
\gamma a_0 + \delta a_1 + db_0 &=& 0, \\
&\vdots& \nonumber \\
\gamma a_{k-1} + \delta a_k + db_{k-1} &=& 0, \\
\gamma a_k + db_k &=& 0.
\end{eqnarray}
The last equation in this chain no longer involves
the differential $\delta$ and can be easily solved.  The
idea, then, is to start the resolution of the cocycle condition
from $a_k$ and to work one's way up until one reaches $a_0$,
which is the quantity of physical interest. [Recall that
$a_0$ defines  a consistent deformation of the Lagrangian.  And
conversely, if $a_0$ is a consistent deformation of the Lagrangian,
then one may complete it by terms of positive antighost number,
as in (\ref{expansion}),
so as to construct a BRST cocycle $a$.  Furthermore, trivial
BRST cocycles (in the cohomological sense) correspond
to trivial deformations (i.e., deformations that can be
absorbed through redefinitions of the field variables)
\cite{BH1,HenneauxCont1}. The reconstruction of the cocycle $a$ from
$a_0$ stops at some antifield number $k$ because $a_0$ is polynomial in
the derivatives (see the argument in \cite{BBH2} section 3).

Before doing this, we shall introduce some useful notations and
give a few solutions.

In the analysis of the BRST cohomology, it turns out that two
combinations of the fields and antifields play a central
r\^ole. The first one combines the field strengths
and the duals of the antifields and is denoted $\tilde H^a $,
\begin{equation}
\tilde H^a = {\overline H}^a + \sum_{j=1}^{p_a+1} {\overline
B}^{*a}_j.
\end{equation}
The second one combines the $p_a$-forms and their associated
ghosts and is denoted $\tilde B^a$,
\begin{equation}
\tB^a = B^a + C^a_1 + \ldots + C^a_{p_a}.
\end{equation}
It is easy to see that both $\tilde H^a$ and $\tB^a$ have
a definite Grassmann parity respectively given by $n-p_a+1$
and $p_a$ modulo $2$. On the other hand, exterior products of $\tilde
H^a$ or
$\tilde B^a$ (including the $\tilde H^a$ and $\tilde B^a$ themselves) are
not homogeneous in form degree and ghost number. 
To isolate a
component of a given form degree $k$ and ghost number $g$, we
enclose the product in brackets $[\ldots]^{k,g}$. The component in
$[A]^{k,g}$ which has definite antighost number $l$ is denoted
$[A]^{k,g}_l$.

Since
products of $\tB^a$ very frequently appear  in the rest of
the analysis, we introduce the following notations,
\begin{equation}\label{conve}
{\cal Q}^{a_1 \ldots a_m}=\tB^{a_1}\ldots \tB^{a_m} \quad
\hbox{and} \quad {\cal Q}^{a_1\ldots
a_m}_{k,g}=[\tB^{a_1}\ldots
\tB^{a_m}]^k_g.
\end{equation}
We shall not write explicitly the wedge product from now on ($dx^0 dx^1$
can clearly only mean $dx^0\wedge dx^1$).

We also define the three
``mixed operators":
$\Delta =
\delta + d$, $\tilde \gamma =\gamma + d$ and $\tilde s=s+d$.

Using those definitions we have the following relations:
\begin{eqnarray}
\Delta {\tilde H}^a &=0, \quad \Delta {\tilde B}^a
 =0,\quad \Delta {H}^a=0\\
\tilde \gamma {\tilde H}^a &=0,  \quad\tilde \gamma
{\tilde B}^a =H^a, \quad\tilde \gamma
{H}^a =0 \label{formulerusse} \\
\tilde s {\tilde H}^a &=0,  \quad \tilde s {\tilde
B}^a =H^a, \quad \tilde s {
H}^a =0.
\end{eqnarray}
Eq. $\tilde \gamma {\tilde
B}^a=H^a$ is known in the literature as the ``horizontality
condition" \cite{Baulieu1}.

It is easy to construct solutions 
of the Wess-Zumino consistency condition out of the 
variables $H^a, \tilde H^a, \tilde B^a$. For example, in 
ghost number zero,
\begin{equation}
    a^{n,0}=[P_b(H^a,\tilde H^a) \tilde B^b]^{n,0},\label{dddd}
\end{equation}
is a solution of (\ref{cocycle}). This can be seen by 
applying $\tilde s$ to $P_b(H^a,\tilde H^a) \tilde B^b$. One gets
$\tilde s(P_b \tilde B^b)=(-)^{\epsilon_P} P_b (\tilde s \tilde
B^b)=(-)^{\epsilon_P} P_b H^b$ and thus, $s[P_b \tilde B^b]^{n,0} +
d[P_b \tilde B^b]^{n-1,1}=[\tilde s (P_b \tilde B^b)]^{n,1}=[P_b
H^b]^{n,1}=0$ (no ghost occurs in $P_b H^b$).  We shall prove in this
article the remarquable property that  all antifield dependent solutions
of the Wess-Zumino consistency condition in ghost number $0$ are in fact
of the form (\ref{dddd}) (modulo antifield independent terms). According
to the discussion at the beginning of Section {\bf 2}, this is equivalent
to proving Theorem {\bf
\ref{central}} since $a^{n,0}_0=[P_b(H^a,\tilde H^a)\tilde B^b]^{n,0}_0=
P_b(H^a,{\overline H^a})B^b$ is of the required form.

\section{Some useful lemmas}
\setcounter{equation}{0}
\setcounter{theorem}{0}
\setcounter{lemma}{0}

In order to construct the general solution of the
(mod $d$) BRST cocycle condition along the lines indicated in the 
previous section, we shall need a few lemmas.

\begin{lemma}
\label{endtrivial}
Let $a_k$ be a solution of $\gamma a_k + db_k =0$, with non-vanishing
antighost number $k$.  Then one has $a_k = a'_k + \gamma m_k
+ dn_k$ where $a'_k$ is annihilated by $\gamma$, $\gamma a'_k = 0$.
\end{lemma}
\proof{The proof proceeds as in the Yang-Mills case: one analyses
the descent equation associated with $\gamma a_k + db_k =0$.  
In \cite{HK2} we have listed all the non-trivial descents 
without taking into account the antifields. However the 
results are unchanged even if one includes the antifields since 
their contributions to non-trivial descents can always be absorbed
by trivial 
terms (the proof of this statement is identical to the one in the 
Yang-Mills case \cite{BBH2}). Therefore, if $a_k$ involves the antifields,
the descent associated with it is necessarily trivial so that one
can find a different representative $a'_k$ in the
same class of $H(\gamma \vert d)$ as $a_k$ which is
annihilated by $\gamma$.}

\begin{lemma}
\label{gammacohomo}
The general solution of $\gamma a_k = 0$ is given by,
\bb
a_k = \sum_I P^I_k \omega^I + \gamma c_k, \label{cocyhg}
\ee
where the $\omega^I$ are polynomials in the undifferentiated "last"
ghosts of ghosts $C^a_{p_a}$ and the $P^I_k$ are spacetime $n$-forms 
with coefficients that are polynomials in the 
field strengths, their derivatives, the antifields and their 
derivatives (these variables will be denoted $\chi$ in the 
sequel).
\end{lemma}
\proof{The proof of this lemma is quite standard. One redefines the
variables into three sets obeying respectively $\gamma x^i=0,\ \gamma
y^\alpha=z^\alpha$,
$\gamma z^\alpha=0$. The
variables $y^\alpha$ and $z^\alpha$ form ``contractible pairs" and
the cohomology is then generated by the (independent) variables $x^i$.
In our case, the $x^i$ are given by $dx^\mu$, the fields
strengths components, the antifields and their
derivatives as well as the
last (undifferentiated) ghosts of ghosts. A complete proof of the lemma in the 
absence of antifields can be found in \cite{HK2}. Here we 
simply note that the 
antifields are automatically part of the $x^i$ variables 
since they are all $\gamma$-closed and do not appear in the 
$\gamma$ variations.}

Using the conventions (\ref{conve}) and dropping the trivial term, we can
write the cocycle (\ref{cocyhg}) as,
$
a_k=\sum_m P_k^{a_1\ldots a_m} [\tB^{a_1} \ldots
\tB^{a_m}]^{0,l}
= \sum_m P_k^{a_1\ldots a_m}
{\cal Q}^{a_1 \ldots a_m}_{0,l},
$
with $l=\sum_m p_{a_m}$.

\begin{lemma}
\label{smallerdegrees}
Let $\alpha$ be an antifield independent $\gamma$-cocycle
that
takes the form
\bb
\alpha = R_1(H^{a_r} , C^{a_r}_{p_{a_r}})
R_2(H^{b_s} , C^{b_s}_{p_{b_s}}), \; p_{b_s}>p_{a_r},
\ee where $R_1$ (respectively $R_2$) is an exterior polynomial in
the curvature form $H^{a_r}$ (respectively $H^{b_s}$) and the
last ghost of ghost $C^{a_r}_{p_{a_r}}$ (respectively
$C^{b_s}_{p_{b_s}}$) such that $p_{b_s}>p_{a_r}$. Assume that
$R_1$ contains no constant term and is trivial  in
$H(\gamma\vert d)$,
\begin{equation}
R_1=\gamma U_1 + dV_1.
\end{equation}
Then, $\alpha$ is also trivial in $H(\gamma
\vert d)$.
\end{lemma}
\proof{This result was proved in 
\cite{HK2}. Since $R_1$ is trivial, it is the obstruction to 
the lift of a $\gamma$-cocycle $\beta_1$ through the descent 
equations of $H(\gamma\vert d)$. Because of the condition 
$p_{b_s}>p_{a_r}$, $\alpha$ then also appears as the 
obstruction to the lift of the $\gamma$-cocycle $\beta_1 R_2$ 
indicating that $\alpha$ is trivial in $H(\gamma\vert d)$.}

The theorem applies in particular when $R_1$ is an
arbitrary polynomial of degree $>0$ in the curvatures
$H^{a_r}$.

\begin{lemma}
\label{triviality}
Let $a$ be a cochain with form-degree $p$ and ghost number $g$, $a\equiv
[a]^{p,g}$, and let $a=a_0+\ldots +a_k$ be its expansion according to
the antighost number, $a_i=[a]^{p,g}_{i}$. Assume that the last term
$a_k$ takes the form $a_k=[P]^{q,-k}_k \chi$ where $P$ is an exterior
polynomial in $\tilde H$ and $H$ and where $\chi \equiv \chi^{p-q,k+g}$
is an exterior polynomial in $H$ and $C^a_{p_a}$ which is trivial in
$H(\gamma\vert d)$, $\chi(H,C)=\gamma m +dn$. Then one can redefine
$a_k$ away by adding $s$-exact terms modulo $d$ to $a$,
\bb 
a=su+dv + \hbox{\ terms of antighost number $< k$}.
\ee
\end{lemma}

\proof{One has $P(\tilde H, H)=[P]^{q-k,0}_0 + \ldots + [P]^{q,-k}_k
+\ldots + [P]^{n,-n+q-k}_{n-q+k}$ and $\tilde s \tilde P=0$. One has
also by assumption, $\chi\equiv \chi^{p-q,k+g}=\gamma m^{p-q,k+g-1}
+dm^{p-q-1,k+g}$ with $m^{p-q,k+g-1}\equiv m$ and $m^{p-q-1,k+g}\equiv
n$. If we define $m^{i,j}\ (i<p-q-1)$ through the descent equation
$\gamma m^{p-q-1,k+g}+ dm^{p-q-2,k+g+1}=0,\ldots$ and $\tilde m =
m^{p-q,k+g-1}+m^{p-q-1,k+g}+m^{p-q-2,k+g+1}+\ldots + m^{0,k+g+p-q-1}$,
one gets, $\chi^{p-q,k+g}=\tilde \gamma \tilde m - dm^{p-q,k+g-1}=\tilde
s \tilde m -dm^{p-q,k+g-1}$. Thus, $\tilde s ((-1)^{\epsilon_P} P\tilde
m)=a_k -P dm^{p-q,k+g-1}$. If we project this equation on the form degree
$p$ of $a_k$, one finds the equation,
\bb
su^{p,g-1}+du^{p-1,g}=a_k-[P]^{q-1,-k+1}_{k-1}dm^{p-q,k+g-1},
\ee
where we have set $u^{p,g-1}\equiv [(-1)^{\epsilon_P}P\tilde m]^{p,g-1}$
and $u^{p-1,g}\equiv [(-1)^{\epsilon_P} P\tilde m]^{p-1,g}$. Thus,
\bb
a_k=su^{p,g-1}+du^{p-1,g}+ \hbox{\ terms of antighost number $<k$},
\ee
which is the desired result.}

\section{Proof of theorem}
\setcounter{equation}{0}
\setcounter{theorem}{0}
\setcounter{lemma}{0}

We now have all the necessary tools required to solve 
the Wess-Zumino consistency condition (\ref{cocycle}).
Consider first the case where the expansion of $a$ (which has total
ghost number $0$) reduces to $a_0$ (no antifields).  Then, $a \equiv
a_0$ fulfills
$\gamma a_0 + db_0 =0$. This equation was investigated in detail in
\cite{HK2}, where it was shown that it has only two types of solutions:
those for which one can assume that $b_0 = 0$, which are the
strictly gauge-invariant terms; and those for which no redefinition
yields $b_0 = 0$ (``semi-invariant terms"), which are exhausted
by the Chern-Simons terms.  Both types
of solutions preserve the form of the gauge symmetries and are in
agreement with the theorem; we  can thus turn to the
case where $a$ involves the antifields, $k \not=0$.

By lemma {\bf \ref{endtrivial}}, one can assume that
the last term $a_k$ in the expansion of $a$ is annihilated
by $\gamma$.  Indeed, the (allowed) redefinition $a \rightarrow
a - sm_k - dn_k$ (see Lemma {\bf {\ref{endtrivial}}}) enables one to do
so.  Then, the next to last equation in the chain (\ref{chain}) implies
$d \gamma b_{k-1}$, i.e., by the algebraic Poincar\'e lemma,
$\gamma b_{k-1} + d c_{k-1} = 0$ for some $c_{k-1}$ (the
cohomology of $d$ is trivial in form-degree $n-1$).

Now, two cases must be considered: either $k>1$, in which case
lemma {\bf \ref{endtrivial}} implies again that one can assume
$\gamma b_{k-1} = 0$ through redefinitions.  Or $k=1$,
in which case $b_{k-1} \equiv b_0$ does not involve the
antifields and may lead to a non trivial
descent.  
This second possibility arises only if $H(\gamma)$
does not vanish in pureghost number one since $a_k\equiv a_1$ must be a 
non-trivial element
of $H^k(\gamma)$ or else can be eliminated through a redefinition.  
In the absence of $1$-forms,
$H^1(\gamma)$ vanishes (lemma {\bf \ref{gammacohomo}}), so we can assume
$k>1$.  The case $k=1$ will be discussed in section {\bf \ref{1forms}}
where we allow for the presence of $1$-forms.

If $k>1$, one can expand the elements $a_k$ and $b_{k-1}$ according to lemma 
{\bf \ref{gammacohomo}},
\bb\label{b0}
a_k = \sum P^I_k \omega^I, \; \; b_{k-1} = \sum
Q^I_{k-1} \omega^I
\ee
($\gamma$-trivial terms can be eliminated).
The next to
last equation in the chain (\ref{chain}) then implies
\bb\label{charcoo}
\delta P^I_k + d Q^I_{k-1} = 0,
\ee
which indicates that $P^I_k$ is a cocycle of the 
cohomology $H(\delta \vert d)$.

This cohomology, which is related to the so-called {\em invariant}
characteristic cohomology, was completely worked out in 
\cite{HKS1}. It was shown that all its representatives can be 
written as the $[\ ]^{n,-k}$ component of an exterior polynomial in $H^a$
and $\tilde H^a$,
\begin{equation}
    P_k^I=[P^I(H^a, \tilde H^a)]^{n,-k}, \quad\quad (k>1).\label{cfgtr}
\end{equation}
It is because of this property that antifield dependent solutions of the
Wess-Zumino consistency condition, which belong a priori to the algebra
generated by all the variables and their individual, successive
derivatives, turn out to be expressible in terms of the forms $H^a$,
$\tilde H^a$ and $B^a$ only. 

Relation (\ref{cfgtr}) implies that the term
$a_k$ of highest antighost number  in the expansion of $a$ is up to
trivial terms of the form,
\bb
a_k = [P^I(H^a, \tilde H^a)]^{n,-k} \omega^I,
\ee
where the pureghost number of the $\omega^I$ must be equal to $k$ in
order to obtain a BRST  cocycle in ghost number $0$.

The question is now: can we construct from the known higher-order
component $a_k$ the components
$a_j$ of lower antighost numbers in order to obtain a 
solution of the Wess-Zumino consistency condition? 

As we 
have seen in Section {\bf \ref{ggen}} this is always possible 
when the $\omega^I$ are linear in the ghosts of ghosts and the 
resulting BRST cocyle is then given by (\ref{dddd}).

We are now going to show that when the $\omega^I$ in $a_k$ are at 
least quadratic in the ghosts of ghosts then one encounters 
an obstruction in the construction of the corresponding 
solution of the Wess-Zumino consistency condition.

To proceed we exhibit explicitly in $a_k$ the $\tilde B^a$ which
correspond to the forms of lowest degree occuring in $a_k$ and
denote them by $\tilde B^{a_i}_1$. The form degree in question is
called $p$. The other $\tilde B^a$ are denoted $\tilde B^{b_j}_2$.
Thus we write
$a_k$ as,
\begin{equation}
a_k=[P_{a_1 \ldots a_r b_1 \ldots b_s}]^{n,-k}[{\tilde
B}^{a_1}_1
\ldots {\tilde B}^{a_r}_1 {\tilde B}^{b_1}_2 \ldots {\tilde
B}^{b_s}_2]^{0,k}.
\end{equation}
Of course, $k>p$ ($a_k$ is at least quadratic in the $\tilde B$). In
fact, $k>p+1$ since there is no $1$-form in the problem.

A direct calculation then shows that the equations $\gamma a_j +\delta
a_{j+1}+db_j=0$ determining $a_{k-1},a_{k-2},\ldots$ have a solution up
to $a_{k-p}$. These solutions are,
\begin{eqnarray}
a_{k-j}&=&[P_{a_1 \ldots a_r b_1 \ldots
b_s}]^{n-j,-k+j}[{\tilde B}^{a_1}_1
\ldots {\tilde B}^{a_r}_1 {\tilde B}^{b_1}_2 \ldots {\tilde
B}^{b_s}_2]^{j,k-j},\label{recu1} \\ && \hspace{7cm} \hbox{for\ 
$0\leq j \leq p$}
\nonumber
\label{recurP}.
\end{eqnarray}
Unless $a_k$ is trivial (i.e., can be removed by the addition of exact
terms to $a$), there is however an obstruction in the construction of
$a_{k-p-1}$. To discuss this obstruction, one needs to know the
ambiguity in the $a_{k-j}$ ($0\leq j\leq p$). One easilly verifies
that it is given by
$a_{k-j}\rightarrow a_{k-j} + m_0 + m_1 +\ldots +m_{j-1}$ where $m_0$
satisfies
$\gamma m_0=0$, $m_1$ satisfies $\gamma m_1 +\delta n_1 + db_1=0,\
\gamma n_1=0$, $m_2$ satisfies $\gamma m_2+\delta n_2+db_2=0,\
\gamma n_2 +\delta l_2+dc_2=0,\ \gamma l_2=0$, etc.
However, none of these
ambiguities except $m_0$ in $a_{k-p}$ can play a role in the
construction of a non-trivial solution. To see this, we note that
$\delta$,
$\gamma$ and $d$ conserve the polynomial degree of the variables
of any given sector\footnote{By sector we mean the variables
corresponding to a given $p$-form and its associated antifields and
ghosts.}. We can therefore work at fixed polynomial degree in the
variables of all the different
$p$-forms. Since
$n_1$, $l_2$, etc. are $\gamma$-closed terms which can be
lifted at least once, they have the generic form $R[H,\tilde
H]{\cal Q}$ where
${\cal Q}$ has to contain a ghost of ghost of degree $p_A<p$.
Because we work at fixed polynomial degree, the presence of such
terms imply that
$P_{a_1\ldots a_r b_1\ldots b_s}$
has to depend on $H^A$ (a dependence on $\tilde H^A$ is not 
possible since by assumption $k>p$). However, $a_k$ is 
then of the form described in Lemma {\bf \ref{triviality}} 
and can be
eliminated from $a$ by the addition of trivial terms and the
redefinition of the terms of antighost numbers $<k$. Therefore we
may now assume that $a_k$ does not contain $H^A$ and that the only
ambiguity in the definitions of the $a_{k-j}$ is $m_0$ in $a_{k-p}$.

Since $k>p$, we have to substitute
$a_{k-p}$ in the equation $\gamma a_{k-p-1}+\delta
a_{k-p}+ db_{k-p-1}=0$. We then get,
\begin{eqnarray}
\gamma a_{k-p-1}+\delta [P_{a_1 \ldots a_r b_1 \ldots
b_s}]^{n-p,-k+p}[{\tilde B}^{a_1}_1
\ldots {\tilde B}^{a_r}_1 {\tilde B}^{b_1}_2 \ldots {\tilde
B}^{b_s}_2]^{p,k-p}\\ +\delta m_0+db_{k-p-1} =0,
\end{eqnarray}
which can be written as,
\begin{eqnarray}
\gamma a^{'}_{k-p-1} + db^{'}_{k-p-1} 
+\delta m_0 \hspace{2cm} \nonumber
\\
 + (-)^{\epsilon_P}r [P_{a_1 \ldots a_r b_1
\ldots b_s}]^{n-p-1,-k+p+1}H^{a_1}_1{\cal Q}^{a_2
\ldots  a_r b_1 \ldots b_s}
_{0,k-p} =0.\label{eqdelobs}
\end{eqnarray}
By acting with $\gamma$ on the above equation we obtain $d\gamma
b^{'}_{k-p-1}=0 \Rightarrow \gamma b^{'}_{k-p-1}+db^{''}_{k-p-1}=0$ 
which means that $b^{'
}_{k-p-1}$ is a $\gamma$ mod $d$ cocycle. Because we have 
excluded $1$-forms from the discussion, $k-p-1>0$ so that 
we may assume that $b^{'}_{k-p-1}$ is strictly
annihilated by $\gamma$. Accordingly,
$db^{'}_{k-p-1}=[d\beta_{a_2\ldots a_r b_1\ldots
b_s}(\chi)]{\cal Q}^{a_2
\ldots  a_r b_1 \ldots b_s}_{0,g+q-p} +\gamma l^{n}_{0,k-p-1}$.
Equation (\ref{eqdelobs}) then reads,
\begin{eqnarray}
&\hspace{-4cm}(-)^{\epsilon_P} r [P_{a_1 \ldots a_r b_1
\ldots b_s}]^{n-p-1,-k+p+1}H^{a_1}_1 \nonumber \\ 
&\hspace{4cm}
+\delta \alpha_{a_2\ldots a_r b_1\ldots b_s}(\chi) + d
\beta_{a_2\ldots a_r b_1\ldots b_s}(\chi)=0,\label{xindeppourP}
\end{eqnarray}
where we have set $m_0=\alpha_{a_2\ldots a_r b_1 \ldots
b_s}(\chi) {\cal Q}^{a_2
\ldots  a_r b_1 \ldots b_s}
_{0,k-p}$. Eq. (\ref{xindeppourP}) implies,
\begin{eqnarray}
[P_{a_1 \ldots a_r b_1
\ldots b_s}]^{n-p-1,-q+p+1}H^{a_1}_1=0,\label{leqasa}
\end{eqnarray}
since $\delta$ and $d$ both increase the number of derivatives of
the $\chi$.
Let us first note that $P_{a_1 \ldots a_r b_1 \ldots b_s}$ cannot
depend on $\tilde H^{c}_1$ because in that case we would have
$k-p-1\leq 0$ which contradicts our assumption that there is no
$1$-form (indeed, the component of form-degree $n$ of a polynomial in
$H^a$ and
$\tilde H^a$ which depends on $\tilde H^c_1$ has maximum antighost number
$p+1$). Therefore,
$P_{a_1 \ldots a_r b_1 \ldots b_s}$ will satisfy 
(\ref{leqasa}) only if it is of the form, $P_{a_1 \ldots a_r b_1
\ldots b_s}=R_{c a_1\ldots a_r b_1\ldots b_s}H_1^{c}$ with $R_{c
a_1\ldots a_r b_1\ldots b_s}$ symmetric in $c \leftrightarrow a_1$
(resp. antisymmetric) if $H_1$ is anticommuting (resp. commuting).
However, using Lemma {\bf \ref{triviality}} we conclude once 
more that in that case
$a_k$ can be absorbed by the addition of trivial terms
and a redefinition of the components of lower antighost number of
$a$. This ends our proof of the statement that for a system 
of $p$-forms with $p \geq 2$ all the antifield dependent 
solutions of the Wess-Zumino consistency conditions in ghost 
number $0$ are of the form (\ref{dddd}).

\section{Presence of $1$-forms}
\label{1forms}
\setcounter{equation}{0}
\setcounter{theorem}{0}
\setcounter{lemma}{0}

If $1$-forms are present in the system of $p$-forms considered, the
solutions in Theorem {\bf \ref{central}} are still valid. However, new
solutions of the Wess-Zumino consistency condition appear,  so the list
is no longer exhaustive. 

The first set
of new solutions, related to the Noether conserved currents of the
theory, arise because
$H^1(\gamma)$ no longer vanishes. Although the term
$b_{k-1}\equiv b_0$ which appears in (\ref{b0}) may lead to a
non-trivial descent, one can show that
(\ref{charcoo}) still holds
\cite{BBH2,theseBK} so that
$P^I\equiv P^a$ has to be an element of $H^n_1(\delta\vert d)$. This
cohomology is isomorphic to the set $a^\Delta$ of non-trivial global
symmetries of the theory. The corresponding solutions of
the Wess-Zumino consistency condition can then be written as,
\bb \label{WZcurrents}
a=k^a_\Delta (j^\Delta B_1^a + a^\Delta C^a_1),
\ee
where the $j^\Delta$ are the Noether currents corresponding to the
$a^\Delta$ and satisfy $\delta a^\Delta + dj^\Delta=0$. The dimension of
this set of solutions is infinite since one can construct infinitely
many conserved currents $j^\Delta$ \cite{HKS1}. This feature is
characteristic of free lagrangians.  Although these solutions define
consistent interactions to first order in the deformation parameter, it
is expected that most of them are obstructed at the second order.
Furthermore, they are severely constrained by Lorentz invariance.

The second set of new solutions of the Wess-Zumino consistency
condition arise because the condition $k-p-1>0$ under (\ref{eqdelobs})
may no longer hold. Indeed, if $p=1$ and $k=2$ then we have $k-p-1=0$. As
above, the term
$b^{'}_{k-p-1}\equiv b^{'}_0$ appearing in (\ref{eqdelobs}) may now lead
to a non-trivial descent in $H(\gamma\vert d)$. According to the
analysis of \cite{HK2}, equation (\ref{leqasa}) is then replaced
by,
\begin{eqnarray}
(-)^{\epsilon_P} r [P_{a_1 \ldots a_r b_1
\ldots b_s}(H^a,\tilde H^a)]^{n-2}_{0}H^{a_1}_1 + V_{a_2\ldots a_r
b_1\ldots b_s}(H^a)=0.
\label{leqasa2}
\end{eqnarray}
The only solution of the above equation for $P^I$ is 
$P^I\equiv k_{abc}{\tilde H}^a_1$ with $k_{abc}$ completely
antisymmetric \cite{BBH2,theseBK}. The corresponding BRST
cocyles are given by,
\bb
a=k_{abc}[\tilde H^a_1 \tilde B^b_1 \tilde B^c_1]^n_0.
\ee
They give rise to the famous Yang-Mills vertex since
$a_0=k_{abc}\overline H^a_1 B^b_1 B^c_1$. 

In particular, the above discussion confirms that
is not possible to construct a Lagrangian with coloured $p$-forms
($p>1$) since vertices of the form $a_0 \sim {\overline H} B A$ (where
$A$ is a $1$-form potential) do not exist. This fact is well appreciated
in the litterature.

\section{Comments and conclusions}

In this paper we have provided the complete proof of the Theorem given
in \cite{HK1} on the consistent deformations of non-chiral free
$p$-forms. The same techniques can be used to study solutions of the
Wess-Zumino consistency condition at other ghost numbers (e.g.,
candidate anomalies) \cite{theseBK}. For instance, one can show that if
all the exterior gauge fields have form degree $\geq 3$, Theorem
{\bf\ref{central}} is also valid for candidate anomalies (the gauge
potential being replaced by the corresponding ghosts of pure ghost number
$1$).

The same methods have also been extended recently to cover chiral
$p$-forms \cite{BeHeSe1}.

\section{Acknowledgements}
This work is suported in part by the ``Actions de
Recherche Concert{\'e}es" of the ``Direction de la Recherche
Scientifique - Communaut{\'e} Fran{\c c}aise de Belgique", by
IISN - Belgium (convention 4.4505.86) and by
Proyectos FONDECYT 1970151 and 7960001 (Chile). 

Bernard Knaepen is supported by a post-doc grant from the
``Wiener-Anspach" foundation.

\end{document}